\documentclass[12pt,epsf,amssymb]{article}
      \def\lsim{\raise0.3ex\hbox{$<$\kern-0.75em\raise-1.1ex\hbox{$\sim$}}}
\def\gsim{\raise0.3ex\hbox{$>$\kern-0.75em\raise-1.1ex\hbox{$\sim$}}}
\def\noi{\noindent}
\def\nn{\nonumber}
\def\bea{\begin{eqnarray}}  \def\eea{\end{eqnarray}}
\def\beq{\begin{equation}}   \def\eeq{\end{equation}}

\def\beeq{\begin{eqnarray}} \def\eeeq{\end{eqnarray}}
\def\R{ {\rm R \kern -.31cm I \kern .15cm}}
\def\C{ {\rm C \kern -.15cm \vrule width.5pt \kern .12cm}}
\def\Z{ {\rm Z \kern -.27cm \angle \kern .02cm}}
\def\N{ {\rm N \kern -.26cm \vrule width.4pt \kern .10cm}}
\def\1{{\rm 1\mskip-4.5mu l} }

\usepackage{graphics}
\usepackage{graphicx}
\usepackage{epsfig}

\begin{document}
\begin{center}
{\large \bf Lower bounds on the curvature of the Isgur-Wise function} \\

\vskip 1 truecm
{\bf A. Le Yaouanc, L. Oliver and J.-C. Raynal}\\
 
{\it Laboratoire de Physique Th\'eorique}\footnote{Unit\'e Mixte de Recherche
UMR 8627 - CNRS }\\    {\it Universit\'e de Paris XI, B\^atiment 210, 91405
Orsay Cedex, France}
\end{center}

\vskip 1 truecm

\begin{abstract}
Using the OPE, we obtain new sum rules in the heavy quark limit of 
QCD, in addition to those previously
formulated. Key elements in  their derivation are the consideration 
of the non-forward amplitude, plus the
systematic use of boundary  conditions that ensure that only a finite 
number of $j^P$  intermediate states (with
their tower of radial excitations) contribute. A study of these sum 
rules  shows that it is possible to bound the
curvature $\sigma^2 =  \xi''(1)$ of the elastic Isgur-Wise function 
$\xi (w)$ in terms of its slope $\rho^2 = - \xi
'(1)$. Besides the bound $\sigma^2 \geq {5 \over 4} \rho^2$, 
previously demonstrated, we find the better bound
$\sigma^2 \geq {1 \over 5} [4 \rho^2 + 3(\rho^2)^2]$. We  show that 
the quadratic term ${3 \over 5} (\rho^2)^2$ has
a  transparent physical interpretation, as it is leading in a 
non-relativistic expansion in the mass of the  light
quark. At the lowest possible value for the slope $\rho^2 = {3  \over 
4}$, both bounds imply the same bound for the
curvature, $\sigma^2 \geq {15 \over 16}$. We point out that these 
results are consistent with the dispersive
bounds, and, furthermore, that they strongly reduce the allowed 
region by the latter for $\xi (w)$. \end{abstract}

\vskip 3 truecm

\noi LPT Orsay 02-108 \par
\noi June 2003\par \vskip 2 truecm

\noindent e-mails : leyaouan@th.u-psud.fr, oliver@th.u-psud.fr
\newpage
\pagestyle{plain}

\section{Introduction}
\hspace*{\parindent}
In a recent paper \cite{1r} we have set a systematic method to obtain
sum rules (SR) in the heavy quark limit of QCD, that relate the
derivatives of the elastic
Isgur-Wise (IW) function $\xi (w)$ to sums over IW functions of
excited states. The method is based on the Operator Product Expansion
(OPE) \cite{2r} applied to heavy
quark transitions \cite{3r} and its key element is the consideration,
following Uraltsev \cite{4r}, of the non-forward amplitude, i.e.
$B(v_i) \to D^{(n)}(v') \to B(v_f)$
with in general $v_i \not= v_f$. Then, the OPE side of the SR
contains the elastic IW function $\xi (w_{if})$ and therefore the SR
depend in general on three variables
$w_i$, $w_f$ and $w_{if}$ that lie within a certain domain. By
differentiation relatively to these variables within the domain and
taking the limit to its boundary,
one finds a very general class of SR that have interesting
consequences on the shape of $\xi (w)$. \par

Let us be more quantitative. As shown in ref. \cite{1r}, using the
OPE -- as formulated for example in \cite{5r} and generalized to $v_i
\not= v_f$ \cite{4r} --, the trace formalism \cite{6r} and
arbitrary heavy quark currents

\beq
\label{1e}
J_1 = \bar{h}_{v'}^{(c)}\ \Gamma_1\ h_{v_i}^{(b)} \quad , \qquad J_2
= \bar{h}_{v_f}^{(b)}\ \Gamma_2\ h_{v'}^{(c)}
\eeq

\noi the following sum rule can be written in the heavy quark limit \cite{1r}~:

\bea
\label{2e}
&&\Big \{ \sum_{D=P,V} \sum_n Tr \left [ \bar{\cal B}_f(v_f)
\Gamma_2 {\cal D}^{(n)}(v') \right ] Tr \left [ \bar{\cal
D}^{(n)}(v') \Gamma_1 {\cal
B}_i(v_i)\right ] \xi^{(n)} (w_i) \xi^{(n)} (w_f) \nn \\
&&+ \ \hbox{Other excited states} \Big \}  = - 2 \xi (w_{if})Tr \left
[ \bar{\cal B}_f(v_f) \Gamma_2 P'_+ \Gamma_1 {\cal
B}_i(v_i)\right ]\ .\eea

In this formula $v'$ is the intermediate meson four-velocity, the projector

\beq
\label{3}
P'_+ = {1 \over 2} (1 + {/ \hskip - 2 truemm v}')
\eeq

\noi comes from the residue of the positive energy
part of the $c$-quark propagator, and $\xi (w_{if})$ is the elastic
Isgur-Wise function that appears because one assumes $v_i \not= v_f$.
${\cal B}_i$ and
${\cal B}_f$ are the $4 \times 4$ matrices of the ground state $B$ or
$B^*$ meson and ${\cal D}^{(n)}$ those of all possible ground state
or excited state $D$
mesons coupled to $B_i$ and $B_f$ through the currents. In formula
(\ref{2e}) we have made explicit the $j = {1 \over 2}^-$ $D$ and
$D^*$ mesons and their radial
excitations. \par

The variables $w_i$, $w_f$ and $w_{if}$ are defined as
    \beq \label{3e}
w_i = v_i \cdot v' \qquad w_f = v_f \cdot v' \qquad w_{if} = v_i \cdot v_f \ .
\eeq

The domain of $(w_i$, $w_f$, $w_{if}$) is \cite{1r}

\beq
\label{4e}
w_i, w_f \geq 1\ , \quad w_i w_f - \sqrt{(w_i^2 - 1) (w_f^2 - 1)}
\leq w_{if} \leq w_i w_f + \sqrt{(w_i^2 - 1) (w_f^2 - 1)}
\eeq

\noi There is a subdomain for $w_i = w_f = w$~:

\beq
\label{5e}
w \geq 1 \qquad 1 \leq w_{if} \leq 2w^2-1 \ .
\eeq

Calling now $L(w_i, w_f, w_{if})$ the l.h.s. and $R(w_i, w_f,
w_{if})$ the r.h.s. of (\ref{2e}), this SR writes

\beq
\label{6e}
L\left ( w_i, w_f, w_{if} \right ) = R \left ( w_i, w_f, w_{if} \right )
\eeq

\noi where $L(w_i, w_f, w_{if})$ is the sum over the intermediate $D$
states and $R(w_i, w_f, w_{if})$ is the OPE side. Within the domain
(\ref{4e}) one can derive
relatively to any of the variables $w_i$, $w_f$ and $w_{if}$

\beq
\label{7e}
{\partial^{p+q+r} L \over \partial w_i^p \partial w_f^q \partial
w_{if}^r} = {\partial^{p+q+r} R \over \partial w_i^p \partial w_f^q
\partial w_{if}^r}
\eeq
\vskip 3 truemm

\noi and obtain different SR taking different limits to the frontiers
of the domain. One must take care in taking these limits, as we point 
out below.\par

Let us parametrize the elastic Isgur-Wise function $\xi (w)$ near zero recoil,

\beq
\label{8e}
\xi (w) = 1 - \rho^2 (w-1) + {\sigma^2 \over 2} (w-1)^2 - \cdots
\eeq

\noi From the SR (\ref{2e}), we gave in ref. \cite{1r} a simple and
straightforward demonstration of both Bjorken \cite{7r} \cite{8r} and
Uraltsev \cite{4r} SR. Both SR
imply the lower bound on the elastic slope

\beq
\label{9}
\rho^2 = - \xi ' (1) \geq {3 \over 4} \ .
\eeq

A crucial simplifying feature of the calculation was to consider, for
the currents (\ref{1e}), vector or axial currents aligned along the
initial and final velocities
$v_i$ and $v_f$. In \cite{1r} we did also obtain, modulo a very
natural phenomenological hypothesis, a new bound on the curvature.

\beq
\label{10}
\sigma^2 = \xi '' (1) \geq {5 \over 4} \rho^2 \geq {15 \over 16} \ .
\eeq

\noi This bound was obtained from the consideration in the SR of the
whole tower of $j^P$ intermediate states \cite{9r}. A crucial feature
of the calculation was the
needed derivation of the projector on the polarization tensors of
particles of arbitrary integer spin \cite{10r}. \par

Using the SR involving the whole sum over all $j^P$ intermediate
states, we pursued our study in \cite{11r} and did demonstrate that
the IW function $\xi (w)$ is an
alternate series in powers of $(w-1)$. Moreover, we did obtain the
bound for the $n$-th derivative at zero recoil $(-1)^n \xi^{(n)}(1)$

\beq
\label{11}
(-1)^n \xi^{(n)} (1) \geq {2n+1 \over 4} (-1)^{n-1} \xi^{(n-1)} \geq
{(2n+1)!! \over 2^{2n}}
\eeq

\noi demonstrating rigorously the bound (\ref{10}) and generalizing
(\ref{9}) and (\ref{10}) to any derivative.\par

The aim of this paper is to investigate whether a systematic
use of the sum rules can allow to obtain better bounds on the
curvature. As we will see below, the answer is positive. The reason
is that only a finite number of
$j^P$ states, with their radial excitations, contribute to the
relevant sum rules and one is left with a relatively simple set of
algebraic linear equations. As we will
see, this is due to the crucial fact that we adopt particular
conditions at the boundary of the domain (\ref{4e}).

\section{Vector and Axial Sum Rules}
\hspace*{\parindent}

   We choose as initial and final states the $B$ meson,

\beq
\label{9e}
{\cal B}_i (v_i) = P_{i+} (- \gamma_5) \qquad {\cal B}_f (v_f) =
P_{f+} (- \gamma_5)
\eeq

\noi where the projectors $P_{i+}$, $P_{f+}$ are defined like in
(\ref{3}). Moreover, we consider vector or axial currents are
projected along the $v_i$ and $v_f$
four-velocities. Choosing the vector currents

\beq
\label{10e}
J_1 = \bar{h}_{v'}^{(c)}\ {/ \hskip - 2 truemm v}_i\ h_{v_i}^{(b)}
\quad , \qquad J_2 = \bar{h}_{v_f}^{(b)}\ {/ \hskip - 2 truemm v}_f\
h_{v'}^{(c)}
\eeq

\noi and gathering the formulas (48) and (89)-(91) of ref. \cite{1r}
we obtain for the SR (\ref{2e}) with the sum of all excited states
$j^P$, as written down in
\cite{11r}~:
$$(w_i + 1) (w_f + 1) \sum_{\ell \geq 0} {\ell + 1 \over 2 \ell + 1}
S_{\ell} (w_i, w_f, w_{if}) \sum_n \tau_{\ell + 1/2}^{(\ell)(n)}(w_i)
\tau_{\ell + 1/2}^{(\ell )(n)}(w_f)$$
\beq
\label{11e}
+ \sum_{\ell \geq 1} S_{\ell} (w_i, w_f, w_{if}) \sum_n \tau_{\ell -
1/2}^{(\ell)(n)}(w_i) \tau_{\ell - 1/2}^{(\ell )(n)}(w_f) = (1 +
w_i+w_f+w_{if}) \xi(w_{if}) \ . \eeq

\vskip 5 truemm
Choosing instead the axial currents

\beq
\label{12e}
J_1 = \bar{h}_{v'}^{(c)}\  {/ \hskip - 2 truemm v}_i \gamma_5 \
h_{v_i}^{(b)} \quad , \qquad J_2 = \bar{h}_{v_f}^{(b)}\ {/ \hskip - 2
truemm v}_f
\gamma_5\ h_{v'}^{(c)}
\eeq

\noi the SR (\ref{2e}) writes, from the formulas (48) and (92)-(94)
of ref. \cite{1r}, obtained in \cite{11r}~:
\bea
\label{13e}
&& \sum_{\ell \geq 0} S_{\ell + 1} (w_i, w_f, w_{if}) \sum_n
\tau_{\ell + 1/2}^{(\ell)(n)}(w_i)
\tau_{\ell + 1/2}^{(\ell )(n)}(w_f)\nn \\
&&+ \ (w_i - 1) (w_f - 1) \sum_{\ell \geq 1} {\ell \over 2 \ell - 1}
S_{\ell - 1} (w_i, w_f, w_{if}) \sum_n \tau_{\ell -
1/2}^{(\ell)(n)}(w_i) \tau_{\ell -
1/2}^{(\ell )(n)}(w_f) \nn \\
&&= - (1 - w_i-w_f+w_{if}) \xi(w_{if}) \ .\eea

Following the formulation of heavy-light states for arbitrary $j^P$
given by Falk \cite{9r}, we have defined in ref. \cite{1r} the IW
functions $\tau_{\ell +
1/2}^{(\ell)(n)}(w)$ and $\tau_{\ell - 1/2}^{(\ell)(n)}(w)$, that
correspond to the orbital angular momentum $\ell$ of the light quark
relative to the heavy
quark, $j = \ell \pm {1 \over 2}$ being the total angular momentum of
the light cloud. For the lower values of $\ell$, one has
the identities with the traditional notation of Isgur and Wise \cite{8r}~:

\beq
\label{14e}
\tau_{1/2}^{(0)}(w) \equiv \xi (w) \quad , \quad \tau_{1/2}^{(1)}(w)
\equiv 2 \tau_{1/2}(w) \quad , \quad \tau_{3/2}^{(1)}(w) \equiv
\sqrt{3}\  \tau_{3/2}(w)
\eeq

\noi where a radial quantum number is implicit. Therefore, the
functions $\tau_{1/2}^{(1)}(w)$ and $\tau_{3/2}^{(1)}(w)$ correspond,
respectively, to the functions
$\zeta (w)$ and $\tau (w)$ defined by Leibovich, Ligeti, Steward and
Wise \cite{12r}. \par

In equations (\ref{10e}) and (\ref{12e}) the quantity $S_n$ is defined by

\beq
\label{15e}
S_n = v_{f\nu_1} \cdots v_{f\nu_n}\ T^{\nu_1 \cdots \nu_n, \mu_1
\cdots \mu_n} \ v_{i\mu_1} \cdots v_{i\mu_n}
\eeq

\noi and the polarization projector $T^{\nu_1 \cdots \nu_k, \mu_1
\cdots \mu_n}$, given by
 
\beq
\label{16e}
T^{\nu_1 \cdots \nu_n, \mu_1 \cdots \mu_n} = \sum_{\lambda}
\varepsilon'^{(\lambda )*\nu_1 \cdots \nu_n} \ \varepsilon'^{(\lambda
)\mu_1 \cdots \mu_n}
\eeq

\noi depends only on the four-velocity $v'$. The tensor $\varepsilon
'^{(\lambda )\mu_1 \cdots \mu_n}$ is the polarization tensor of a
particle of integer spin $J =
n$, symmetric, traceless, i.e. $\varepsilon '^{(\lambda )\mu_1 \cdots
\mu_n} g_{\mu_i\mu_j} = 0$ ($i\not= j \leq n$), and transverse to
$v'$, $v'_{\mu_i} \varepsilon
'^{(\lambda )\mu_1 \cdots \mu_n} = 0$ ($i \leq n$) \cite{1r} \cite{10r}.\par

Moreover, as demonstrated in the Appendix A of ref. \cite{1r}, $S_n$ is
given by the following expression~:

\beq
\label{17e}
S_n(w_i,w_f,w_{if}) = \sum_{0 \leq k \leq {n \over 2}} C_{n,k} (w_i^2
- 1)^k (w_f^2 - 1)^k (w_i w_f - w_{if})^{n-2k}
\eeq

\noi with

\beq
\label{18e}
C_{n,k} = (-1)^k {(n!)^2 \over (2n) !} \ {(2n - 2k) ! \over k! (n-k)
! (n-2k)!} \ .
\eeq

The relation

   \beq \label{19e}
\left . L^V(w_i, w_f, w_{if}) \right |_{w_{if} = 1, w_i = w_f = w} =
\left . R^V(w_i, w_f, w_{if}) \right |_{w_{if} = 1, w_i = w_f = w}
\eeq

\noi gives, dividing by $2(w+1)$, Bjorken SR \cite{7r} \cite{8r},
including now the whole sum of intermediate states~:
\bea
\label{20e}
&&{w + 1 \over 2} \sum_{\ell \geq 0} {\ell + 1\over 2 \ell + 1}
C_{\ell}(w^2 - 1)^{\ell} \sum_n \tau_{\ell + 1/2}^{(\ell)(n)}(w)
\tau_{\ell + 1/2}^{(\ell )(n)}(w)\nn \\
&&+ \ {w - 1 \over 2} \sum_{\ell \geq 1}  C_{\ell}(w^2 - 1)^{\ell-1}
\sum_n \tau_{\ell - 1/2}^{(\ell)(n)}(w)
\tau_{\ell - 1/2}^{(\ell )(n)}(w) = 1
\eea

\noi where, from (\ref{17e}) and (\ref{18e})

\beq
\label{21e}
S_n (w, w, 1) = C_n(w^2 - 1)^n \qquad C_n = \sum_{0 \leq k \leq {n
\over 2}} C_{n,k} = 2^n {(n!)^2 \over (2n)!} \ .
\eeq

\noi Remember that, usually, the first terms in the sum (\ref{20e})
are written in the notation (\ref{14e}) of Isgur and Wise \cite{8r}~:

\beq
\label{newequation26}
{w + 1 \over 2} \sum_n \left [ \xi^{(n)}(w)\right ]^2 + (w-1) \sum_n
\left \{ 2 \left [ \tau_{1/2}^{(n)}(w) \right ]^2 + (w+1)^2 \left [
\tau_{3/2}^{(n)}(w) \right ]^2
\right \} + \cdots = 1 \ . \eeq

Going now to the axial current SR (\ref{13e}), the condition

\beq \label{32e}
\left . L^A(w_i, w_f, w_{if}) \right |_{w_{if} = 1, w_i = w_f = w} =
\left . R^A(w_i, w_f, w_{if}) \right |_{w_{if} = 1, w_i = w_f = w}
\eeq

\vskip 5 truemm

\noi gives again, dividing this time by $2(w-1)$, the complete
Bjorken SR (\ref{20e}). Notice that, as it must, one obtains the same
SR from the vector (\ref{10e})
and the axial current (\ref{12e}) because, from (\ref{21e}), one has

   \beq
\label{33e}
(2 n + 1) C_{n+1} = (n+1) C_n \ .
\eeq

\section{Equations from the Vector Sum Rule}
\hspace*{\parindent}
In what follows, to look for independent relations, we make use of
the fact that the SR (\ref{11e}) and (\ref{13e}) are symmetric in the
exchange $w_i \leftrightarrow
w_f$.\par

Let us first consider the derivatives of the SR for vector currents
(\ref{11e}) relatively to $w_{if}$ with the boundary condition 
$w_{if} = 1$. For $w_{if} = 1$,
the domain (\ref{4e}) implies~:

\beq
\label{neqn}
w_i = w_f = w \ .
\eeq

\noi We define therefore

\bea
\label{neqn1}
&&\left . L_V(w_{if}, w) \equiv L_V(w_{if}, w_i, w_f) \right |_{w_i = 
w_f = w} \nn \\
&&\left . R_V(w_{if}, w) \equiv R_V(w_{if}, w_i, w_f) \right |_{w_i = 
w_f = w} \ .
\eea

\noi We then take the $p+q$ derivatives

\beq
\label{26new}
\left ( {\partial^{p+q} L_V \over \partial w_{if}^p \partial w^q } 
\right )_{w_{if}=w=1} = \left (
{\partial^{p+q} R_V \over \partial w_{if}^p
\partial w^q} \right )_{w_{if}=w=1} \eeq

\noi and exploit systematically the obtained relations. To get 
information on the curvature $\sigma^2$ of the
elastic IW function  (\ref{8e}) we need to go up to the second order 
derivatives. As we will see, a crucial feature
of the  adopted boundary condition is that only a finite number of 
$j^P$  states, with their tower of radial
excitations, contribute to the sum rule. Notice that we could have 
derived first relatively to $w_i$ and take the
limit $w_i = 1$, and then derive with respect to $w = w_{if} = w_f$. 
We do not obtain however new information from these sum rules that 
with
the former boundary conditions. \par

Let us proceed with care and begin with the first order derivatives.
  From (\ref{11e}) and (\ref{26new}), we obtain the following results.
\par

With the notation (\ref{26new}), for $p = q = 0$ we obtain the 
trivial result $\xi (1) = \xi (1)$, while for the
derivatives $p = 1$,  $q = 0$ we obtain Bjorken SR for the slope $\rho^2$~:

\beq
\label{28new}
\rho^2 = {1 \over 4} + {2 \over 3} \sum_n \left [
\tau_{3/2}^{(1)(n)}(1) \right ]^2 + {1 \over 4} \sum_n \left [
\tau_{1/2}^{(1)(n)}(1) \right ]^2 \ .
\eeq

\noi The relation to the Isgur-Wise notation is given by (\ref{14e}). \par

Going now to the second order derivatives, we find the following
relations. For $p = 0$, $q = 1$ we get $\xi (1) = \xi (1)$. For a 
purpose that will appear clear below, we make
explicit the IW functions between $j^P =
{1 \over 2}^-$ states using the notation of Isgur and Wise
$\xi^{(n)}(w)$ (\ref{14e}). \par

For $p = 2$, $q = 0$~:

\beq
\label{29new}
\rho^2 - 2\sigma^2 +  {12 \over 5} \sum_n \left
[\tau_{5/2}^{(2)(n)}(1)\right ]^2 + \sum_n \left
[\tau_{3/2}^{(2)(n)}(1)\right ]^2= 0
\eeq

$p = 1$, $q = 1$~:

\bea
\label{30new}
&&\rho^2 - {4 \over 3} \sum_n \left [ \tau_{3/2}^{(1)(n)}(1) \right ]^2 \nn \\
&&- \ {8 \over 3} \sum_n \tau_{3/2}^{(1)(n)}(1)
\tau_{3/2}^{(1)(n)'}(1) - \sum_n \tau_{1/2}^{(1)(n)}(1)
\tau_{1/2}^{(1)(n)'}(1) \nn \\
&&- \ 2 \sum_n \left [ \tau_{3/2}^{(2)(n)}(1) \right ]^2 -  {24 \over
5} \sum_n \left [ \tau_{5/2}^{(2)(n)}(1) \right ]^2 = 0
\eea

$p = 0$, $q = 2$~:

\bea
\label{32new}
&&1 - 8\rho^2 + 4 \sigma^2 + \ 4 \sum_n \left [ \xi^{(n)'}(1) \right ]^2 + 8
  \sum_n \left [ \tau_{3/2}^{(1)(n)}(1) \right ]^2 + \sum_n
\left [ \tau_{1/2}^{(1)(n)}(1) \right
]^2\nn \\ &&+ \ {32 \over 3} \sum_n \tau_{3/2}^{(1)(n)}(1)
\tau_{3/2}^{(1)(n)'}(1) +4 \sum_n \tau_{1/2}^{(1)(n)}(1)
\tau_{1/2}^{(1)(n)'}(1) \nn \\
&&+ \ {8 \over 3} \sum_n \left [ \tau_{3/2}^{(2)(n)}(1) \right ]^2 +
{32 \over 5} \sum_n\left [ \tau_{5/2}^{(2)(n)}(1) \right ]^2  = 0 \ .
\eea

\vskip 5 truemm

The equations (\ref{29new})-(\ref{32new}) are a set of linear
equations in the elastic slope $\rho^2$ and the curvature $\sigma^2$,
and the following quantities, that are
series on the radial excitations, indicated by the sums over $n$~:

\bea
\label{1eqn}
&&\sum_n \left [ \xi^{(n)'}(1) \right ]^2 \\
\label{2eqn}
&&\sum_n \left [ \tau_{3/2}^{(1)(n)}(1) \right ]^2 \\
\label{3eqn}
&&\sum_n \left [ \tau_{1/2}^{(1)(n)}(1) \right ]^2 \\
\label{4eqn}
&&- \sum_n  \tau_{3/2}^{(1)(n)}(1) \ \tau_{3/2}^{(1)(n)'}(1) \\
\label{5eqn}
&&- \sum_n  \tau_{1/2}^{(1)(n)}(1) \ \tau_{1/2}^{(1)(n)'}(1) \\
\label{6eqn}
&&\sum_n \left [ \tau_{3/2}^{(2)(n)}(1) \right ]^2 \\
\label{7eqn}
&&\sum_n \left [ \tau_{5/2}^{(2)(n)}(1) \right ]^2
\eea

\noi We realize first that, due to the fact that we compute the
second derivatives in (\ref{26new}) $(p + q = 2)$ and use the
boundary conditions $w_{if} = w = 1$, the series in $j^P$ states is 
truncated and includes at
most the $\ell = 2$ states $j^P = {3 \over 2}^-, {5\over 2}^-$,
corresponding to the unknowns
(\ref{6eqn}) and (\ref{7eqn}). On the other hand (\ref{1eqn}) is the
square of the derivatives at zero recoil of the lowest $j^P = {1\over
2}^-$, and (\ref{2eqn}) and
(\ref{3eqn}) depend on the IW functions of the transitions to the
$P$-wave states $j^P = {1 \over 2}^+, {3 \over 2}^+$, that are simply
related to the slope $\rho^2$
through Bjorken and Uraltsev SR, as we write down below again.
Finally, we have two other unknowns (\ref{4eqn}) and (\ref{5eqn})
that involve the derivatives of the
$P$-wave IW functions $\tau_{3/2}^{(1)(n)}(w)$,
$\tau_{1/2}^{(1)(n)}(w)$ at zero recoil. These quantities were
already introduced in ref. \cite{1r}.

\section{Equations from the Axial Sum Rule}
\hspace*{\parindent}
Let us now consider likewise the derivatives of the SR for axial
currents (\ref{13e}) with the boundary condition $w_{if}
= 1$, $w_i = w_f = w \to 1$~:

\beq
\label{33new}
\left ( {\partial^{p+q} L_A \over \partial w_{if}^p \partial w^q}
\right )_{w_{if}=w=1} = \left (
{\partial^{p+q} R_A \over \partial w_{if}^p
\partial w^q}\right )_{w_{if}=w=1} \ . \eeq

\vskip 5 truemm

\noi Since all terms in (\ref{13e}) vanish for $w_i = w_f = w_{if} =
1$, to obtain information on the curvature $\sigma^2$ we will need to
go up to the third order
derivatives. \par

With the notation (\ref{33new}), for $p = q = 0$ we get $0 = 0$,
and for $p=1$, $q=0$ and $p = 0$, $q = 1$, $\xi (1) = \xi (1)$.
  \par

For the second order derivatives we obtain the following results. For
$p=2$, $q=0$ and $p=q=1$ we get the same relation

\beq
\label{35new}
\rho^2 = \sum_n \left [ \tau_{3/2}^{(1)(n)}(1) \right ]^2
\eeq

\noi while for $p=0$, $q=2$, we get Bjorken SR~:

\beq
\label{36new}
\rho^2 = {1 \over 4} + {2 \over 3} \sum_n \left [
\tau_{3/2}^{(1)(n)}(1) \right ]^2 + {1 \over 4} \sum_n \left [
\tau_{1/2}^{(1)(n)}(1) \right ]^2 \ .
\eeq

\noi Both equations (\ref{35new}) and (\ref{36new}) imply Uraltsev SR

\beq
\label{37new}
{1 \over 3} \sum_n \left [ \tau_{3/2}^{(1)(n)}(1) \right ]^2 - {1
\over 4} \sum_n \left [ \tau_{1/2}^{(1)(n)}(1) \right ]^2 = {1 \over
4}
\eeq

\vskip 5 truemm

\noindent using the notation (\ref{14e}). \par

Going now to the third order derivatives, we obtain the following results. \par

For $p = 3$, $q = 0$~:

\beq
\label{38new}
\sigma^2 =  2 \sum_n \left [\tau_{5/2}^{(2)(n)}(1)\right ]^2
\eeq

$p = 2$, $q = 1$~:

\beq
\label{39new}
\sigma^2 =  2 \sum_n \tau_{3/2}^{(1)(n)}(1) \tau_{3/2}^{(1)(n)'}(1) +
6 \sum_n \left [ \tau_{5/2}^{(2)(n)}(1) \right ]^2
\eeq

$p = 1$, $q =  2$~:

\bea
\label{41new}
&&\sigma^2 + \sum_n  \left [ \xi^{(n)'}(1) \right ]^2 +  2 \sum_n \left [
\tau_{3/2}^{(1)(n)}(1) \right ]^2  \nn \\
&&+ \ 8 \sum_n \tau_{3/2}^{(1)(n)}(1) \tau_{3/2}^{(1)(n)'}(1)  + {2
\over 3}\sum_n\left [ \tau_{3/2}^{(2)(n)}(1) \right ]^2 + {48 \over
5} \sum_n \left [ \tau_{5/2}^{(2)(n)}(1) \right ]^2= 0
\eea

$p=0$, $q=3$~:

\bea
\label{43new}
&&- 3 \rho^2 + 3\sigma^2  + 3 \sum_n  \left [ \xi^{(n)'}(1)
\right ]^2  + 4 \sum_n \left [ \tau_{3/2}^{(1)(n)}(1)
\right ]^2 \nn \\
&&+ \ 8 \sum_n \tau_{3/2}^{(1)(n)}(1)
\tau_{3/2}^{(1)(n)'}(1) + 3 \sum_n \tau_{1/2}^{(1)(n)}(1)
\tau_{1/2}^{(1)(n)'}(1) \nn \\
&&+ \ 2 \sum_n \left [ \tau_{3/2}^{(2)(n)}(1) \right ]^2 +
{24 \over 5} \sum_n \left [ \tau_{5/2}^{(2)(n)}(1) \right ]^2 = 0
   \ .  \eea

\noi The equations (\ref{38new})-(\ref{43new}) depend on $\rho^2$,
$\sigma^2$ and the same set of unknowns listed in
(\ref{1eqn})-(\ref{7eqn}). Let us now look for the set
of linearly independent equations that can be obtained from the sets
of equations obtained from the Vector and Axial SR. \section{Linearly
independent relations}
\hspace*{\parindent}
Let us concentrate on the equations (\ref{29new})-(\ref{32new}) and
(\ref{38new})-(\ref{43new}) obtained respectively from the Vector and
Axial Sum Rules. \par

Using Bjorken SR

\beq
\label{44new}
\rho^2 = {1 \over 4} + {2 \over 3} \sum_n \left [
\tau_{3/2}^{(1)(n)}(1) \right ]^2 + {1 \over 4} \sum_n \left [
\tau_{1/2}^{(1)(n)}(1) \right ]^2 \ ,
\eeq

\noi the relation

\beq
\label{45new}
   {4 \over 3} \rho^2 - 1 = \sum_n \left [ \tau_{1/2}^{(1)(n)}(1) \right ]^2
\eeq

\noi obtained from (\ref{28new}) and (\ref{35new}), and (\ref{35new})
and (\ref{38new}) we obtain finally the following set of relations
that
are linearly independent

\beq
\label{54new}
\rho^2 = - \ {4 \over 5} \sum_n \tau_{3/2}^{(1)(n)}(1)
\tau_{3/2}^{(1)(n)'}(1) + {3 \over 5} \sum_n \tau_{1/2}^{(1)(n)}(1)
\tau_{1/2}^{(1)(n)'}(1)
\eeq

\beq
\label{55new}
\sigma^2  = - \sum_n \tau_{3/2}^{(1)(n)}(1) \tau_{3/2}^{(1)(n)'}(1)
\eeq

\beq
\label{56new}
\sigma^2  = 2 \sum_n \left [ \tau_{5/2}^{(2)(n)}(1) \right ]^2
\eeq

\beq
\label{57new}
\rho^2  - {4 \over 5} \sigma^2 + \sum_n \left [
\tau_{3/2}^{(2)(n)}(1) \right ]^2 = 0
\eeq

\beq
\label{58new}
{4 \over 3} \rho^2  - {5 \over 3} \sigma^2 + \sum_n \left [
\xi^{(n)'}(1) \right ]^2  = 0 \ .
\eeq

Relations (\ref{54new}) and (\ref{55new}) were obtained in ref.
\cite{1r}, and relation (\ref{56new}) and (\ref{57new}) in \cite{11r}.
The systematic study of the present paper using all possibilities
(\ref{26new}) and (\ref{33new}) involving the curvature gives the new
equation (\ref{58new}).

\section{Bounds on the curvature}
\hspace*{\parindent}
The last two equations (\ref{57new}) and (\ref{58new}) involve the
curvature with a negative sign and positive definite quantities.
Making explicit in the sum
$\sum\limits_n [\xi^{(n)'}(1)]^2$ the ground state IW function slope
$\xi^{(0)'}(1) = - \rho^2$, one obtains the two equations

\beq
\label{60}
\rho^2 - {4 \over 5} \sigma^2 + \sum_n |\tau_{3/2}^{(2)(n)}(1)|^2 = 0 \ .
\eeq

\beq
\label{59}
{4 \over 3} \rho^2 + (\rho^2)^2 - {5 \over 3} \sigma^2 + \sum_{n\not=
0} |\xi^{(n)'}(1)|^2 = 0
\eeq

\noi that imply respectively the bounds~:
   \bea
\label{57e}
&&\sigma^2 \geq {5 \over 4} \ \rho^2 \\
&&\sigma^2 \geq {1 \over 5} \left [ 4 \rho^2 + 3(\rho^2)^2 \right ] \ .
\label{58e}
\eea

\noi The bound (\ref{57e}) was obtained in ref. \cite{1r} using the
relations (\ref{54new}) and (\ref{55new}) and making the assumption

\beq
\label{59e}
- \sum_n \tau_{1/2}^{(1)(n)}(1) \ \tau_{1/2}^{(1)(n)'}(1) \geq 0 \ .
\eeq

\noi Later, (\ref{57e}) was demonstrated rigorously in ref.
\cite{11r} and generalized to the $n$-th derivative. However, in this
latter paper, only
derivatives relatively to $w_{if}$ were taken, while in the present
work a systematic use of (\ref{26new}) and (\ref{33new}) is carried
out. \par

The inequality (\ref{58e}) is the best of the bounds that we have
obtained for $\sigma^2$ for any value of $\rho^2$, and is the main
result of this paper. \par

Interestingly enough, both bounds (\ref{57e}) and (\ref{58e})
coincide at the lower bound $\rho^2 \geq {3 \over 4}$ implied by
Bjorken and Uraltsev SR, (\ref{28new}) and
(\ref{37new}). At the value $\rho^2 = {3 \over 4}$ one then gets
indeed the same absolute bound (i.e., independent of $\rho^2$) for
$\sigma^2$, namely
(\ref{10}) $\sigma^2 \geq {15 \over 16}$.

\section{Implication on the P-wave IW functions at zero recoil}
\hspace*{\parindent}
Let us now express the sums of products of the $P$-wave Isgur-Wise
functions ${1 \over 2}^- \to {1 \over 2}^+$ and their derivatives
$\sum\limits_n
\tau_{3/2}^{(1)(n)}(1) \tau_{3/2}^{(1)(n)'}(1)$ and $\sum\limits_n
\tau_{1/2}^{(1)(n)}(1) \tau_{1/2}^{(1)(n)'}(1)$ in terms of $\rho^2$
and $\sigma^2$. From
(\ref{54new}) and (\ref{55new}) we obtain, using now the notation of
Isgur and Wise (\ref{14e})~:

\beq
\label{68new}
-\sum_n \tau_{3/2}^{(n)}(1) \tau_{3/2}^{(n)'}(1) = {1 \over 3} \sigma^2
\eeq

\beq
\label{69new}
-\sum_n \tau_{1/2}^{(n)}(1) \tau_{1/2}^{(n)'}(1) = - {5 \over 12}
\rho^2 + {1 \over 3} \sigma^2 \ .
\eeq
\vskip 5 truemm

\noi Using the bounds (\ref{9}) and (\ref{10}) for $\rho^2$ and
$\sigma^2$ one finds

\bea
\label{70new}
&&- \sum_n \tau_{3/2}^{(n)}(1) \tau_{3/2}^{(n)'}(1) \geq {5 \over 16} \\
&& -\sum_n \tau_{1/2}^{(n)}(1) \tau_{1/2}^{(n)'}(1) \geq 0 \ .
\label{71new}
\eea

Strictly speaking, these relations do not give information
on the slope of the lowest $n = 0$ IW functions
$\tau_{3/2}^{(0)'}(1)$ and $\tau_{1/2}^{(0)'}(1)$.
However, if the $n = 0$ state dominates the sum, the inequalities
(\ref{70new}) and (\ref{71new}) imply that the slopes
$\tau_{3/2}^{(0)'}(1)$ and $\tau_{1/2}^{(0)'}(1)$
are negative, as it is plausible on physical grounds for form factors
that do not involve radially excited states. \par

This is indeed the case for the Bakamjian-Thomas type of quark
models, that satisfy IW scaling \cite{13r} and Bjorken and Uraltsev
sum rules \cite{14r}. We have
conjectured in \cite{1r} that this class of models presumably satisfy
all the SR of the heavy quark limit of QCD. \par

In the Bakamjian-Thomas model one finds for the phenomenologically
successful spectroscopic model of Godfrey and Isgur \cite{15r}, the
numbers

\beq
\label{72new}
- \tau_{3/2}^{(0)}(1) \tau_{3/2}^{(0)'}(1) = 0.43
\eeq

\beq
\label{73new}
- \tau_{1/2}^{(0)}(1) \tau_{1/2}^{(0)'}(1) = 0.04
\eeq

\noi that by themselves satisfy the preceding bounds, so that the $n
= 0$ state seems to give a dominant contribution to the l.h.s. of
(\ref{70new}) and (\ref{71new}).
 
\section{Non-relativistic limit of the bounds}
\hspace*{\parindent}
There is a simple intuitive argument to understand the term ${3 \over
5} (\rho^2)^2$ in the best bound (\ref{58e}). Let us consider the
non-relativistic quark
model, i.e. a non-relativistic light quark $q$ interacting with a
heavy quark $Q$ through a potential. The form factor -- to be
identified with the IW function --
has then the simple form~:

\beq
\label{67e}
F({\bf k}^2)= \int d {\bf r} \ \varphi^+_0(r) \exp \left ( i {m_q
\over m_q + m_Q} {\bf k} \cdot {\bf r} \right ) \varphi_0 (r)
\eeq

\noi where $\varphi_0(r)$ is the ground state radial wave function.
In the small momentum transfer limit, the IW variable $w$ writes, in
the initial heavy hadron
rest frame~:

\beq
\label{68e}
w \cong 1 + {{\bf v}'^2 \over 2} = 1 + {{\bf k}^2 \over 2m_Q^2} \ .
\eeq

\noi Identifying the non-relativistic IW function $\xi_{NR}(w)$ with
the form factor $F({\bf k}^2)$ (\ref{67e}), one finds, because of
rotational invariance~:

\beq
\label{69e}
\xi_{NR}(w) \cong 1 - m_q^2 <0|z^2|0> (w-1) + {1 \over 2}\ {1 \over
3} \ m_q^4 <0|z^4|0>(w-1)^2 + \cdots
\eeq

\noi where $|0>$ stands for the ground state wave function, and we have
neglected in the $(w-1)^2$ coefficient subleading terms in powers of
$1/(m_q z)$ (internal velocity). Therefore, one has the following
expressions for the slope and the curvature, in the non-relativistic
limit~:

\beq
\label{70e}
\rho_{NR}^2 = m_q^2<0|z^2|0> \qquad \sigma^2_{NR} \cong {1 \over 3} \
m_q^4 <0|z^4|0> \ .
\eeq

\noi From spherical symmetry one has

\beq
\label{71e}
<0|z^4|0> \ = {1 \over 5} <0|r^4|0> \ .
\eeq

\noi Using now completeness $\sum\limits_n |n><n| = 1$,

\beq
\label{72e}
<0|r^4|0> \ = |<0|r^2|0>|^2 + \sum_{n \not= 0} |<n|r^2|0> |^2
\eeq

\noi we use again spherical symmetry

\beq
\label{73e}
<0|r^4|0> \ = 9 |<0|z^2|0>|^2 + 9 \ \sum_{n \not= 0, rad} \ |<n|z^2|0>|^2
\eeq

\noi where the latter sum runs only over radial excitations. \par

Therefore, from (\ref{70e})-(\ref{73e}) we can rewrite
$\sigma_{NR}^2$ under the form

\beq
\label{74e}
\sigma^2_{NR} = {3 \over 5} \left \{ \left [ m_q^2|<0|z^2|0>|\right
]^2 + m_q^4 \sum_{n \not= 0, rad} \ |<n|z^2|0>|^2 \right \}
\eeq

\noi or

\beq
\label{75e}
\sigma_{NR}^2  = {3 \over 5} \left  [ \rho^2_{NR} \right ]^2 + {3
\over 5} \ m_q^4\ \sum_{n \not= 0, rad} \ |<n|z^2|0>|^2
\eeq

\noi and therefore

\beq
\label{76e}
\sigma_{NR}^2 \geq {3 \over 5} \ \left  [ \rho^2_{NR} \right ]^2 \ .
\eeq

Notice that, denoting by $R$ the bound state radius and $m_q$ the
light quark mass, in the non-relativistic limit, just from
expressions (\ref{70e}), one can
see that $\rho^2_{NR}$ scales like $m_q^2R^2$, while $\sigma^2_{NR}$
scales like $m_q^4R^4$ and both the l.h.s. and the r.h.s. in
(\ref{76e}) scale in the same way. \par

Going back to the relativistic bounds (\ref{57e})-(\ref{58e}), we
observe that the terms proportional to $\rho^2$ are subleading in the
non-relativistic expansion and correspond to  relativistic corrections
specific to QCD in the heavy quark limit. In the non-relativistic limit
$\rho^2 \sim m_q^2R^2  \gg 1$, and the power $(\rho^2)^2$ is leading.
We can understand therefore the appearance of the term ${3 \over 5}
(\rho^2)^2$ in the r.h.s. of the inequality (\ref{58e}).\par

\section{An example of fit to the data}
\hspace*{\parindent}
An interesting phenomenological remark is that the simple
parametrization for the IW function \cite{15r}

\beq
\label{77e}
\xi (w) = \left ( {2 \over w + 1} \right )^{2 \rho^2}
\eeq

\noi gives

\beq
\label{78e}
\sigma^2 = {\rho^2 \over 2} + (\rho^2)^2
\eeq

\noi that satisfies the inequalities (\ref{57e})-(\ref{58e}) if
$\rho^2 \geq {3 \over 4}$, i.e. for all values allowed for $\rho^2$.
Moreover, interestingly, at the
lowest bound of the slope $\rho^2 = {3 \over 4}$, (\ref{78e}) implies
precisely the lower bound of the curvature $\sigma^2 = {15 \over
16}$, as pointed out in \cite{11r}.
\par

Notice that, in ref. \cite{15r}, within the class of Bakamjian-Thomas
quark models, the approximate form (\ref{77e}) was found with $\rho^2
= 1.02$ in the particular case
of the spectroscopic model of Godfrey and Isgur. This gives a
curvature (\ref{78e}) $\sigma^2 = 1.55$, close to the bound
(\ref{58e}), that gives $\sigma^2 \geq 1.44$,
stronger than the bound (\ref{57e}), that implies $\sigma^2
\geq 1.27$. \par

As a simple example of a fit with the simple function (\ref{77e}), we
can use BELLE data on $\bar{B}^0 \to D^{*+} e^- \bar{\nu}$ for the
product $|V_{cb}|{\cal F}^*(w)$ \cite{16r},
as shown in the Figure. \par

The function ${\cal F}^*(w)$ is equal to the Isgur-Wise function $\xi
(w)$ in the heavy quark limit. Assuming only departures of this limit
at $w = 1$, i.e. fitting $\xi (w)$ from the data with

\beq \label{nlleq}
|V_{cb}| {\cal F}^*(w) = |V_{cb}| {\cal F}^*(1) \xi (w)
\eeq

\noi we obtain the following
results for the normalization and
the slope~:

\beq
\label{nlleq1}
{\cal F}^*(1) |V_{cb}| = 0.036 \pm 0.002 \qquad \rho^2 = 1.15 \pm 0.18
\eeq

\noi with the other derivatives of $\xi (w)$ fixed by (\ref{77e}) 
(Fig. 1). \par

\begin{figure}
\begin{center}
\epsfig{file=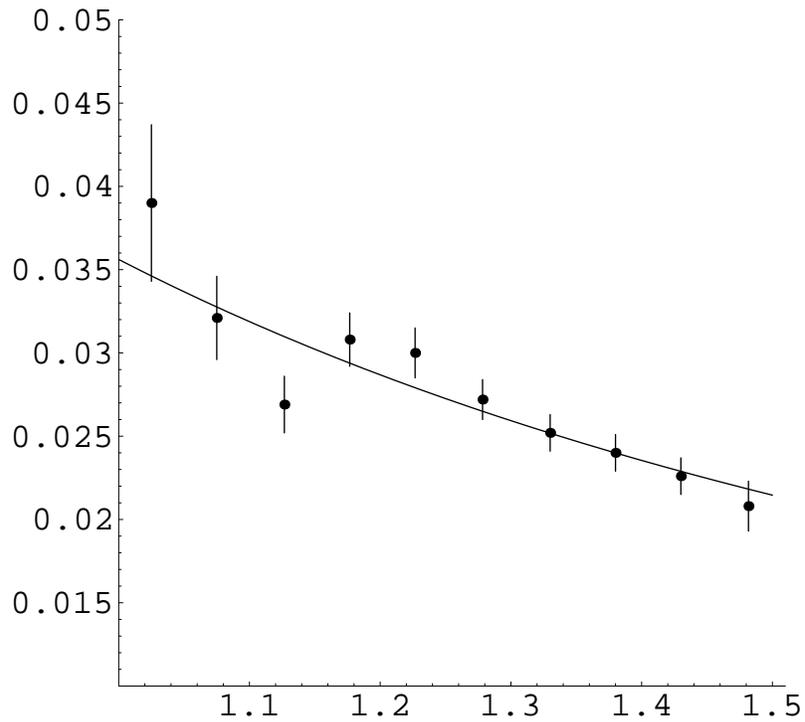, height=12.cm}
\caption{\label{fig1}\footnotesize{\sl Fit to ${\cal F}^*(w)|V_{cb}|$ using the
phenomenological formula (\ref{77e}) and the BELLE data for $\overline{B} \to
D^*\ell \nu$ [16], assuming only violations to the heavy quark limit at
$w= 1$. The fit gives the results (\ref{nlleq1}).}} \end{center} \end{figure}

As we can see, the determination of ${\cal F}^*(1)|V_{cb}|$ is rather precise,
while $\rho^2$ has a larger
error. However, the values obtained for $|V_{cb}|$ and $\rho^2$ are
strongly correlated. It is
important to point out that the most precise data points are the ones
at large $w$, so that higher derivatives contribute importantly in
this region. Due to the alternate
character of $\xi (w)$ as a series of $(w-1)$, one does not see
strongly the curvature of $\xi (w)$ in the Figure, but the curve is
definitely not close to a straight
line. Linear fits, as are commonly used, should be ruled out at the
view of the bounds that we have found. \par

We must emphasize that the fit that we present is a simple exercise
in the heavy quark limit. Radiative corrrections and $1/m_Q$
corrections, that enter in the relation
between the actual function $F(w)$ and its heavy quark limit $\xi
(w)$,  should be taken into account, although this does not seem to
be an easy task \cite{17r}.

\section{Comparison with the dispersive bounds}
\hspace*{\parindent}
A considerable effort has been developed to formulate dispersive
constraints on the shape of the form factors in $\bar{B} \to D^*\ell
\nu$ \cite{18r}-\cite{22r}. The
starting point are the analyticity properties of two-point functions
and positivity of the corresponding spectral functions. Then,
dispersion relations relate the
hadronic spectral functions to the QCD two-point functions in the
deep Euclidean region, and positivity allows to bounds the
contribution of the relevant states, leading to
constraints on the semileptonic form factors.\par

We will now compare our method, that gives information on the
derivatives of the Isgur-Wise function, with the dispersive approach.
\par

A first remark to be made is that our approach, based on Bjorken-like
SR, holds {\it in the physical region} of the semileptonic decays
$\bar{B} \to D^{(*)}\ell \nu$ and
{\it in the heavy quark limit}. However, concerning this last
simplifying feature, we should underline that there is no objection
of principle to include in the
calculation radiative corrections and subleading corrections in
powers of $1/m_Q$. \par

The dispersive approach starts from bounds {\it in the crossed
channel} by comparison of the OPE and the sum over hadrons coupled to
the corresponding current,
$\bar{B}\bar{D}$, $\bar{B}\bar{D}^*$, $\cdots$ Then, one analytically
continues to the physical region of the semileptonic decays. This is
done for a single reference
form factor, for example the combination

\beq
\label{equation1}
V_1(w) = h_+(w) - {m_B - m_D \over m_B + m_D} \ h_-(w)
\eeq

\noi that enters in the $\bar{B} \to D\ell \nu$ rate. In the heavy
quark limit $h_-(w) = 0$, $V_1(w) = h_+(w) = \xi (w)$. Ratios of the
remaining form factors to $V_1(w)$
are computed {\it in the physical region} by introducing $1/m_Q$ and
$\alpha_s$ corrections to the heavy quark limit. The dispersive
approach considers {\it
physical quark masses}, in contrast with the heavy quark limit of our
method. \par

The two approaches are quite different in spirit and in their
results. However, it can be interesting to compare numerically our
bounds with the ones of the dispersive
approach, as they happen to be complementary. We must however keep in
mind precisely the differences between the two methods.\par

We have demonstrated in \cite{11r} that the IW function $\xi (w)$ is
an alternating series in powers of $(w - 1)$, with the moduli of the
derivatives satisfying the bounds
(\ref{11}) and (\ref{58e}).

\subsection{Comparison with the work of Caprini, Lellouch and Neubert}
\hspace*{\parindent}
Let us consider the main results of ref. \cite{21r}, that are
summarized by the one-parameter formula

\beq
\label{equation2}
{V_1(w) \over V_1(1)} \cong 1 - 8 \rho^2 z + (51\rho^2 - 10)z^2 -
(252 \rho^2 - 84) z^3
\eeq

\noi with the variable $z(w)$ defined by

\beq
\label{equation3}
z = {\sqrt{w+1} - \sqrt{2}\over \sqrt{w+1} + \sqrt{2}}
\eeq

\noi and the allowed range for $\rho^2$ being

\beq
\label{equation4}
- 0.17 < \rho^2 < 1.51 \ .
\eeq

\noi Of course, the function ${V_1(w) \over V_1(1)}$ contains finite
mass corrections that are absent at present in our method.
Nevertheless, let us first compare
these results with our lower bounds (\ref{11}), assuming the rough
approximation

\beq
\label{equation5}
{V_1(w) \over V_1(1)} \cong \xi (w) \ .
\eeq

\noi Of course, since the expansion (\ref{equation2}) stops at third
order in $z$, it would only make sense in the comparison to go up to
the third derivative of $\xi (w)$.
Using our notation, the results of Section 4 of ref. \cite{21r} for
the first derivatives write, from the expansion (\ref{equation2}), in
terms of the slope $\rho^2$~:

\bea
\label{(1)}
&&\xi '' (1) = {1 \over 32} (67 \rho^2 - 10) \\
&&\xi '''(1) = - {1 \over 256} (1487 \rho^2 - 372)
\label{(2)}
\eea

\noi with $\rho^2$ in the range (\ref{equation4}). From (\ref{(1)})
and (\ref{(2)}), using the notation $\xi (w) = 1 - \rho^2(w-1) +
c(w-1)^2 + d(w-1)^3 + \cdots$
one gets the numerical relations \cite{21r}~: $c \cong 1.05 \rho^2 -
0.15$, $d \cong - 0.97 \rho^2 + 0.24$.\par

   The lower bound (\ref{equation4}) on $\rho^2$ is very weak~:
there is a region for $\rho^2$ that is below the Bjorken bound
$\rho^2 \geq 1/4$, and this reflects on the regions allowed for the
higher derivatives. Of course, we must
always keep in mind that in these considerations we are neglecting
finite mass and $\alpha_s$ corrections. Moreover, one should notice
that the series is not alternate in
the whole range (\ref{equation4}). We have verified
analytically that the expression (\ref{equation2}) is an alternate
series in
powers of $(w-1)$ for values of $\rho^2 \geq {1 \over 4}$, i.e. for
$\rho^2$ satisfying precisely Bjorken bound. It is not clear to us
whether this is just a numerical
coincidence or it has a deeper significance, i.e. if it extends to
the complete series in powers of $z$.\par

Let us now comment on the implications of our bounds (\ref{11}). The
first important remark is that, within the simplifying hypothesis
(\ref{equation5}), the range
(\ref{equation4}) is considerably tightened by the lower bound on
$\rho^2 \geq {3 \over 4}$ implied by Bjorken and Uraltsev sum rules.
Therefore, we will consider
hereafter, instead of (\ref{equation4}), the improved range

\beq
\label{(3)}
{3 \over 4} \leq \rho^2 < 1.51
\eeq

\noi that shows that our type of lower bounds are complementary to the
upper bounds obtained from dispersive methods. Within the hypothesis of
the heavy quark limit, the region allowed by the dispersive bounds for
$\xi (w)$ with $\rho^2$ within the range (\ref{equation4}) is obviously much
reduced by the bounds (\ref{(3)}) (Fig. 2).\par

\begin{figure}[h!]
\begin{center}
\epsfig{file=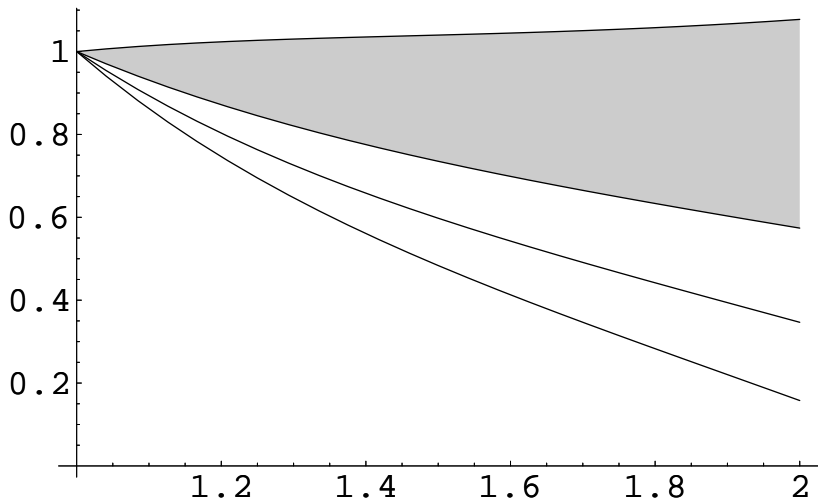, height=9.5cm}
\caption{\label{fig2}\footnotesize{\sl The upper (lower) curves are 
the representations of $\xi (w)$
according to the dispersive approach of Caprini et al [21] 
(\ref{equation2})-(\ref{equation5}).
The upper (lower) curve correspond to $\rho^2 = - 0.17$ ($\rho^2 =
1.51)$. The shadowed region is the region forbidden by the Uraltsev
bound $\rho^2 \geq {3 \over 4}$. The remaining allowed region
corresponds to (\ref{(3)}). The curve within this allowed region is the best
fit to BELLE data [16], normalized to $w = 1$, that gives ${\cal
F}^*(1)|V_{cb}| = 0.036 \pm 0.002$, $\rho^2 = 1.16 \pm 0.15$, in practice
the same fit as (\ref{nlleq1}) with the phenomenological formula (\ref{77e}).}}
\end{center} \end{figure}

On the other hand, it follows that the second and third derivatives
(\ref{(1)}), (\ref{(2)}) satisfy the bounds (\ref{11})

\bea
\label{(4)}
&&\xi '' (1) = {1 \over 32} (67 \rho^2 - 10 \geq {5 \over 4} \rho^2 \\
&&- \xi ''' (1) = {1 \over 256} (1487 \rho^2 - 372) \geq {7 \over 4}
\xi '' (1) = {7 \over 4} \ {1 \over 32} (67 \rho^2 - 10)
\label{(5)}
\eea

\noi respectively for $\rho^2 \ \gsim \ 0.36$ and $\rho^2 \ \gsim \
0.42$. Therefore, for values of $\rho^2$ that are within the range
(\ref{(3)}), these inequalities are
{\it a fortiori} satisfied, but they are not satisfied in the whole
range (\ref{equation4}). \par

Finally, let us look for the implications of our improved bound on
the curvature, eq. (\ref{58e}). Combining the linear dependence
obtained from dispersive methods
(\ref{(1)}) with the inequality (\ref{58e}) one obtains the condition

\beq
\label{(6)}
{1 \over 32} (67 \rho^2 - 10) \geq {1 \over 5} \left [ 4 \rho^2 +
3(\rho^2)^2 \right ]
\eeq

\noi that gives the range

\beq
\label{(7)}
0.28 \ \lsim \ \rho^2 \ \lsim \ 1.88 \ .
\eeq

\noi Interestingly, the condition (\ref{(6)}) gives {\it by
itself} an upper bound for $\rho^2$ that is of the same order than
the upper bound (\ref{equation4}).
Moreover, the range (\ref{(7)}) contains the improved range
(\ref{(3)}), and things appear to be coherent.

\subsection{Comparison with the work of Boyd, Grinstein and Lebed}
\hspace*{\parindent}
Let us now compare with the dispersive method results of the work of
Boyd, Grinstein and Lebed \cite{22r}. In this work, the QCD part of
the calculation includes
$\alpha_s$ and non-perturbative (condensate) corrections, and new
poles below the annihilation threshold ignored in \cite{21r}. In a
form that allows to make the
comparison with our results, the authors of \cite{22r} obtain the
following expansion for the scalar form factor

\bea
\label{(1e)}
\widetilde{f}_0(w) &=& \widetilde{f}_0(1) + \left [ 1.72a_1 - 0.77
\widetilde{f}_0(1)\right ] (w-1) \nn \\
&&+ \left [ - 1.74a_1 + 0.21a_2 + 0.55\widetilde{f}_0(1)\right ]
(w-1)^2 + \cdots
\eea

\noi where $\widetilde{f}_0(w)$ has been defined by Caprini and
Neubert \cite{20r}

\beq
\label{(2e)}
\widetilde{f}_0(w(q^2)) = {f_0(q^2)\over (M_B - M_D)  \sqrt{M_B M_D} (w+1)}
\eeq

\noi with

\beq
\label{(3e)}
f_0(q^2) = \left ( M_B^2 - M_D^2 \right ) f_+(q^2) + q^2 f_-(q^2)
\eeq

\noi $f_{\pm}(q^2)$ being the form factors governing the rate
$\overline{B} \to D \ell \nu$ ($q= p-p'$)~:

\beq
\label{(4e)}
<D(p') |V_{\mu}|B(p)> = f_+(q^2) (p + p')_{\mu} + f_-(q^2)(p-p')_{\mu}\ .
\eeq

\noi Heavy quark symmetry implies

\beq
\label{(5e)}
\widetilde{f}_0(w(q^2)) \cong \xi (w) \ .
\eeq

\noi The coefficients $a_n$ in (\ref{(1e)}) are defined by the
expression for a generic form factor \cite{21r}, \cite{22r}

\beq
\label{(6e)}
F(z) = {1 \over P(z) \varphi (z)} \sum_{n=0}^{\infty} a_n z^n
\eeq

\noi where $z$ is defined by (\ref{equation3}). The functions $P(z)$
and $\varphi (z)$ -- respectively the Blaschke factor and the outer
function -- contain the
subthreshold singularities in the annihilation channel, respectively
the $B_c$ poles and the kinematic singularities. The basic result of
the dispersive approach is that the
coefficients $a_n$ of the series obey

\beq
\label{(7e)}
\sum_{n=0}^{\infty} a_n^2 \leq 1 \ .
\eeq

To compare with our bounds we proceed like we did above. Since in the
heavy quark limit (\ref{(5e)}) holds, we set $\widetilde{f}_0(1) = 1$
and write the Isgur-Wise
function in terms of the coefficients $a_n$~:

\bea
\label{(8e)}
\xi(w) &\cong& 1 + (1.72a_1 - 0.77) (w-1) \nn \\
&&+ \left (-1.74 a_1 + 0.21a_2 + 0.55\right )(w-1)^2 + \cdots
\eea

\noi Notice that in (\ref{(1e)}) and (\ref{(8e)}) it does not make
much sense to consider higher powers $(w-1)^n$ ($n \geq 3$) unless
the corresponding $a_n$ ($n \geq 3$)
are introduced. Then, our lower bounds (\ref{11}) write

\bea
\label{(9e)}
&&- 1.72a_1 + 0.77 \geq {3 \over 4} \nn \\
&&2\left ( - 1.74a_1 + 0.21a_2 + 0.55 \right ) \geq {5 \over 4} (-
1.72a_1 + 0.77)
\eea

\noi implying, respectively

\bea
\label{(10e)}
&&a_1 \leq 0.01 \nn \\
&&a_2 \geq 3.17 a_1 - 0.33  \ .
\eea

\noi Since, from (\ref{(10e)}) and (\ref{(7e)}) we have

\beq
\label{(11e)}
- 1 \leq a_1 \leq 0.01
\eeq

\noi and the coefficient of $a_1$ in (\ref{(10e)}) is large, the whole range

\beq
\label{(12e)}
- 1 \leq a_2 \leq 1
\eeq

\noi is allowed. This seems to support the statement of ref. \cite{22r} that
$a_2$ cannot always be neglected.\par

Moreover, using now the quadratic bound (\ref{58e}), one obtains

\beq
\label{(13e)}
3(\rho^2)^2 - 6 \rho^2 + 2(1 - a_2) \ \lsim\ 0
\eeq

\noi and therefore

\beq
\label{(15e)}
- 0.5 \ \lsim \ a_2 \leq 1
\eeq

\noi giving the following range for $\rho^2$ in terms of $a_2$~:

\beq
\label{(14e)}
1 - \sqrt{{1 + 2a_2 \over 3}} \ \lsim\ \rho^2 \ \lsim\ 1 + \sqrt{{1 +
2a_2 \over 3}}
\eeq

\noi giving the wide range

\beq
\label{112}
0 \ \lsim \ \rho^2 \ \lsim \ 2 \ .
\eeq

\noi  For $a_2 = 0$, implicitely assumed in ref. \cite{21r}, one finds
the range

\beq
\label{(16e)}
0.42 \ \lsim \ \rho^2 \ \lsim \ 1.58
\eeq

\noi a domain qualitatively consistent but somewhat narrower than the
corresponding one (\ref{(7)}) obtained from the linear relation
between the curvature and the slope
given by ref. \cite{21r}. \par

In conclusion, there is no contradiction between the dispersive
bounds and the type of bounds that we have obtained using
Bjorken-like sum rules in the heavy quark
limit. The latter appear rather as lower bounds that are
complementary to the upper bounds of the dispersive approach,
tightening considerably the allowed range for
$\rho^2$ and for the higher derivatives of $\xi (w)$ as well.

\section{A phenomenological Ansatz for the Isgur-Wise function and the
dispersive constraints} \hspace*{\parindent} At the light of the
preceding discussion, we are now going to address the question of
whether the phenomenological Ansatz for the IW function proposed in
Section 9,

\beq \label{1ref} \xi (w) = \left ( {2 \over w+1}\right )^{2\rho^2}
\eeq

\noi satisfies, assuming the heavy quark limit (\ref{equation5}) or
(\ref{(5e)}), the constraints of the dispersive approach. \par

We will follow here the formulation of Boyd et al. \cite{22r} and
consider the form factors $f_+(q^2)$ and $f_0(q^2)$ defined by
(\ref{(3e)}) and (\ref{(4e)}). In the heavy quark limit, one has the
relations

\bea \label{2ref} &&f_+(q^2(w)) \cong {M_B + M_D \over 2 \sqrt{M_BM_D}}
\ \xi (w) \\ &&f_0(q^2(w)) \cong (M_B - M_D) \sqrt{M_BM_D} (w+1) \xi
(w) \ . \label{3ref} \eea

We denote now generically any of these form factors by $F(q^2(w))$, or
through the transformation (\ref{equation3}), $F(q^2(z))$. \par

We adopt the phenomenological formula (\ref{1ref}) for $\xi (w)$ and
define the corresponding series (\ref{(6e)})

\beq \label{4ref} \sum_{n=0}^{\infty} a_n\ z^n = P(z) \ \varphi (z)\
F(z) \eeq

\noi where $P(z)$ and $\varphi (z)$ are the Blaschke factor and the
outer function of the corresponding form factors. \par

We want now to compare the coefficients $a_n$ obtained from
(\ref{2ref})-(\ref{3ref}), assuming $F(z) = \xi (w(z))$ given by 
(\ref{1ref}), to the condition (\ref{(7e)})

\beq \label{5ref} \sum_{n=0}^{\infty} a_n^2 \leq 1 \ . \eeq

The outer functions $\varphi (z)$ and the Blaschke factors $P(z)$ for
$f_+(q^2)$ and $f_0(q^2)$ are given in ref. \cite{22r}, respectively by
formula (4.23) and Table 1 and by formula (4.25) and Table 3. We have
singled out $f_+(q^2)$ and $f_0(q^2)$ as given by (\ref{2ref}) and
(\ref{3ref}) but we could have taken any other form factor related, up
to kinematic factors, to $\xi (w)$. Of course, the results for the
coefficients $a_n$ would differ according to the considered form
factor. \par

We use the numerical parameters of this paper, and two choices for 
$\rho^2$ in formula (\ref{1ref}), namely $\rho^2 = 1.023$, that
corresponds to the Isgur-Wise function obtained within the
Bakamjian-Thomas scheme from the Godfrey-Isgur spectroscopic model, as
found in ref. \cite{15r}, and $\rho^2 = 1.15$ obtained from the fit of
Section 9. \par

Denoting the Blaschke factor
and outer function for each form factor by the corresponding
subindices, we find, for $\rho^2 = 1.023$,  the series (\ref{4ref}) 
for $f_+(q^2(z))$~:

\bea \label{6ref} P_+(z) \varphi_+(z) f_+ (q^2(z)) &=& 0.0143 - 0.0179z
- 0.1164z^2 + 0.3277z^3 \nn \\ &&- \ 0.1995 z^4 - 0.4497 z^5 + 1.2347z^6
+ \cdots \eea

\noi and for $f_0(q^2(z))$~:

\bea \label{7ref} P_0(z) \varphi_0(z) f_0 (q^2(z)) &=& 0.0834 -
0.1750z - 0.1725z^2 + 0.8673z^3 \nn \\ &&- \ 1.1600 z^4 - 0.8943 z^5 -
0.4346 z^6 + \cdots \eea

For $\rho^2 = 1.15$ we find, respectively~:

\begin{eqnarray*}
P_+(z) \varphi_+(z) f_+(q^2(z)) &=& 0.0143 - 0.0326z - 0.0907 z^2 + 
0.4294z^3 -\\
&&  - \ 0.5779z^4 - 0.0306z^5 + 1.3868 z^6 + \cdots
\end{eqnarray*}

\noi and

\begin{eqnarray*}
P_0(z) \varphi_0(z) f_0(q^2(z)) &=& 0.0834 - 0.2599z + 0.0484z^2 + 
0.9094z^3 -\\
&& - \ 2.0079z^4 + 2.5089z^5 - 2.3596z^6 + \cdots
\end{eqnarray*}

Comparing to the condition (\ref{5ref}), we observe two points. First,
the first coefficients have squares well below 1, specially for
$f_+(q^2)$. That this happens to be the case for this form factor, that
has three Blaschke factors, reinforces the idea that one should be
closer to the IW function, according to our hypothesis
(\ref{2ref})-(\ref{3ref}), as the number of Blaschke factors increases.
Second, high powers of $z$ have coefficients that can be of $O(1)$ and
they are strongly dependent on the value of $\rho^2$, specially for
$f_0(q^2)$, that has only two Blaschke factors. We have actually
observed that the behaviour of the coefficients oscillate, as can be
seen in (\ref{7ref}).\par

Our conclusion is that, owing to the fact that the coefficients, up to
order $z^3$ included, satisfy the condition (\ref{5ref}), the
``dipole'' formula (\ref{1ref}) gives, on phenomenological grounds, a good
enough representation of the form factors (\ref{2ref}), (\ref{3ref}).

\section{Conclusions} \hspace*{\parindent} In conclusion, using sum
rules in the heavy quark limit of QCD, as formulated in ref.
\cite{1r}, we
have found an improved bound for the curvature of the Isgur-Wise
function $\sigma^2 = \xi '' (1) \geq {1 \over 5} [ 4 \rho^2 +
3(\rho^2)^2 ]$ that implies the already
demonstrated \cite{1r} \cite{11r} absolute bound $\sigma^2 \geq {15
\over 16}$.\par

Beyond the simple Ansatz (\ref{77e}) introduced above, any
phenomenological parame\-trization of $\xi (w)$ intending to fit the
CKM matrix element $|V_{cb}|$ in $B \to
D^{(*)}\ell \nu$ should have, for a given slope $\rho^2$ satisfying
the bound (\ref{9}), a curvature $\sigma^2$ satisfying the new bound
(\ref{58e}). \par

Moreover, we discuss these bounds in comparison with the dispersive
approach. We show that there is no contradiction, our bounds
restraining the region for $\xi (w)$ allowed by this latter method.\par

\vskip 0.5 truecm

\noi {\large \bf Acknowledgements} \par

We acknowledge support from the EC contract HPRN-CT-2002-00311 
(EURIDICE), and Zoltan Ligeti for a useful remark.

\end{document}